\documentstyle[12pt]{article}
\input psfig.tex
\input tex.def

\def\title#1{\relax\vspace*{2cm}{\large{\bf #1}}\par\vspace*{13.5pt}}
\def\author#1{{#1}\par\vspace*{13.5pt}}
\def\affil#1{{\it #1}\par}
\def\abstract{\vspace*{27pt}ABSTRACT\par\relax}
\def\section#1{\par{#1}\par}
\def\subsection#1{\par\underline{#1}\par}
\def\subsubsection#1{\par\underline{#1.}\ \ }

\def\gtorder{\mathrel{\raise.3ex\hbox{$>$}\mkern-14mu
             \lower0.6ex\hbox{$\sim$}}}
\def\ltorder{\mathrel{\raise.3ex\hbox{$<$}\mkern-14mu
             \lower0.6ex\hbox{$\sim$}}}
\def\ltsima{$\; \buildrel < \over \sim \;$}
\def\simlt{\lower.5ex\hbox{\ltsima}}
\def\gtsima{$\; \buildrel > \over \sim \;$}
\def\simgt{\lower.5ex\hbox{\gtsima}}
\def\hst{{\it HST }}
\def\Ha{H$\alpha$ }

\def\acknow{\par ACKNOWLEDGMENTS\par}
\newenvironment{references}{\section{REFERENCES}\vspace*{.5cm}%
\parindent=0pt\frenchspacing%
\parskip=1pt plus 1pt minus 1pt%
\interlinepenalty=1000\tolerance=400%
\pretolerance=10000\hyphenpenalty=10000%
\everypar={\hangindent=1.6pc}
}{}
 
\begin{document}

\title{THE ROLE OF STARS IN THE ENERGETICS OF LINERs}

\author{Dan Maoz}
\affil{School of Physics and Astronomy, Tel-Aviv University, Tel-Aviv 69978, Israel}

\footnotesize 
Invited review
to appear in {\it The 32$^{nd}$ COSPAR Meeting, The AGN-Galaxy Connection}, 
ed. H.~R. Schmitt, A.~L. Kinney, and L.~C. Ho 
(Advances in Space Research).
\normalsize

\abstract
Imaging studies have shown that $\sim 25\%$ of LINER galaxies
display a compact nuclear UV source.
I compare the \hst ultraviolet (1150--3200~\AA) spectra that are
now available for seven 
such ``UV-bright'' LINERs.
 The spectra of NGC 404, NGC 4569, and NGC 5055 show clear
absorption-line
signatures of massive stars, indicating a stellar origin for the UV 
continuum.
Similar features are probably present in NGC 6500.
The same stellar signatures {\it may} be present but undetectable in NGC 4594,
due to the low signal-to-noise ratio of the spectrum, and in M81 and NGC 4579, due
to superposed strong, broad emission lines.
The compact
central UV continuum source that is observed in these galaxies is
a nuclear star cluster rather than a low-luminosity active galactic
nucleus (AGN), at least in some cases. 
At least four of the LINERs suffer from an ionizing photon
deficit, in the sense that the ionizing photon flux inferred from the
observed far-UV continuum is insufficient to drive the optical H~I 
recombination lines.
Examination of the nuclear X-ray flux of each galaxy shows a high
X-ray/UV ratio  in the four ``UV-photon starved''
LINERs. In these four objects, a separate component, emitting predominantly in the
extreme-UV, is the likely ionizing agent, and is
perhaps unrelated to the observed nuclear UV emission.
Future observations can determine whether the UV continuum in
LINERs is always dominated by a starburst or, alternatively,
that there are two types of UV-bright LINERs: starburst-dominated
and AGN-dominated. Interestingly, recent results show that starbursts
dominate the nuclear energetics in many Seyfert 2s as well.

\section{INTRODUCTION}

Low-ionization
nuclear emission-line regions (LINERs) are detected in the
nuclei of a large fraction of all bright nearby galaxies
(Ho, Filippenko, \& Sargent 1997a; Ho, this volume).
Since their definition as a class by Heckman (1980), they have elicited debate
as to their nature and relation, if any, to active galactic
nuclei (AGNs). On the one hand, the luminosities of most LINERs
are unimpressive compared to ``classical'' AGNs, and can easily be produced 
by processes other than accretion onto
massive black holes. Indeed, LINER-like spectra are sometimes
seen to arise in ``non-nuclear'' environments, such as cooling flows.
On the other hand, a variety of observables
point to similarities and continuities between AGNs
and at least some LINERs (see Ho, this volume). 
If LINERs represent the low-luminosity end of the AGN phenomenon,
then they are the nearest and most common examples, and their 
study is germane to understanding AGN demographics, quasar evolution,
dormant black holes in quiescent galaxies, the X-ray background,
and the connection between AGN-like and starburst-like activity.

The ultraviolet (UV) sensitivity and angular resolution of the {\it Hubble Space Telescope}
(\hst) is providing new clues toward understanding LINERs.
Maoz et al. (1996a) carried out a UV (2300 \AA) imaging survey of the 
central regions of 110 nearby galaxies with the Faint Object Camera (FOC)
on {\it HST}. As reported in Maoz et al. (1995), five among the 25 LINERs
in their sample revealed nuclear UV sources, in most cases unresolved,
implying physical sizes $\ltorder 2$ pc. Maoz et al. (1995) argued
that the UV sources in these ``UV-bright'' LINERs could be the
extension of the ionizing continuum, which is rarely seen in LINERs
at optical wavelengths due to the strong background from the normal
bulge population. The compactness of the sources suggested that they could be 
nonstellar in nature, although compact star clusters of such
luminosity were also possible. A similar fraction of UV-bright
nuclei was found in a 2200 \AA\ imaging survey of 20 LINER and
low-luminosity Seyfert 2 galaxies carried out with the WFPC2 camera on
\hst by
Barth et al. (1996a; 1998). A compact, isolated, nuclear UV source
has also been found in FOC images of the LINER NGC 4594 at $\sim 3400$ \AA\ 
 by Crane et al. (1993), and in  WFPC2 images of the LINER M81 in a broad
(1100 \AA\ to 2100 \AA) bandpass by Devereux, Ford, \& Jacoby (1997).
  The UV-bright LINERs were obvious
targets for follow-up spectroscopy with \hst. Faint Object Spectrograph
(FOS) observations have
been analyzed for the LINERs M81 (Ho, Filippenko, \& Sargent 1996), NGC 4579
(Barth et al. 1996b), NGC 6500 (Barth et al. 1997), and NGC 4594
(Nicholson et al. 1998).

The spectra of M81 and NGC 4579, which in the optical range have
weak broad wings in the \Ha line (Filippenko \& Sargent 1985; 1988;
Ho et al. 1997b), are reminiscent of AGNs  in the UV, with strong, broad emission
lines superposed on a featureless
continuum. 
As discussed by Barth et al. (1996b), in NGC 4579
the FOS 2200 \AA\ flux is less than 1/3 that implied by the FOC
measurement made 19 months earlier, implying variability of a nonstellar
continuum source.
It would be valuable, however, to confirm such
variability in this and other LINERs. 
Ho et al. (1996) and Barth et al. (1996b) concluded that
these two LINERs are most probably AGNs. NGC~6500 and NGC 4594, on the other hand,
have only weak and narrow emission lines on top of a UV continuum.
Based on the
resolved appearance of the UV source in the \hst WFPC2 F218W image,
and the tentative detection of optical Wolf-Rayet features, Barth et al. (1997)
concluded that the UV emission in NGC~6500 is likely dominated by 
massive stars, though contribution from a scattered AGN
component could not be excluded. For NGC 4594, Nicholson et al. (1998) favor
an AGN interpretation.
Maoz et al. (1998) presented \hst spectra for three additional 
UV-bright LINERs (NGC 404, NGC 4569, and NGC 5055), and compared the
properties of all seven UV-bright LINERs observed spectroscopically
with \hst to date. I summarize below the results of that analysis.

\section{FAR-UV SPECTRAL SIGNATURES OF MASSIVE STARS}

Figure 1 shows most of the G130H spectrum of all
seven LINERs.
 The objects are ordered with the two broad-lined
LINERs  (M81 and NGC 4579) on top, and the other LINERs in order
of decreasing $f_{\lambda}$. Figure 2 shows in more detail
the same spectral
region for  NGC 4569, NGC 404,
NGC 5055, and NGC 6500.
Overlayed on each spectrum
(thin line) is a 
scaled, normalized version of the \hst GHRS spectrum of the 
``B'' clump in NGC 1741, a starburst galaxy, which
is described by Conti, Leitherer \& Vacca (1996).
The broad, blueshifted absorption
 profiles of C IV $\lambda1549$ and Si IV $\lambda1400$
 (and also N V $\lambda 1240$) in NGC 1741-B
are the signatures of winds produced by massive
stars (see Leitherer, Robert, \& Heckman 1995). 
The other, narrower, 
absorption lines are of both photospheric and interstellar-medium (ISM) origin.
The characteristic starburst features seen in the spectrum of
 NGC 1741-B are also observed in star-forming regions in 30 Doradus
(Vacca et al. 1995) and NGC 4214 (Leitherer et al. 1996), in 
star-forming galaxies at redshifts $z\approx 2-3$ (Steidel et al. 1996;
 Lowenthal et al. 1997),
and in the $z=3.8$ starburst radio galaxy 4C 41.17 (Dey et al. 1997).

\begin{figure}
\vbox{
\vbox{
\hskip 0.1truein
\psfig{file=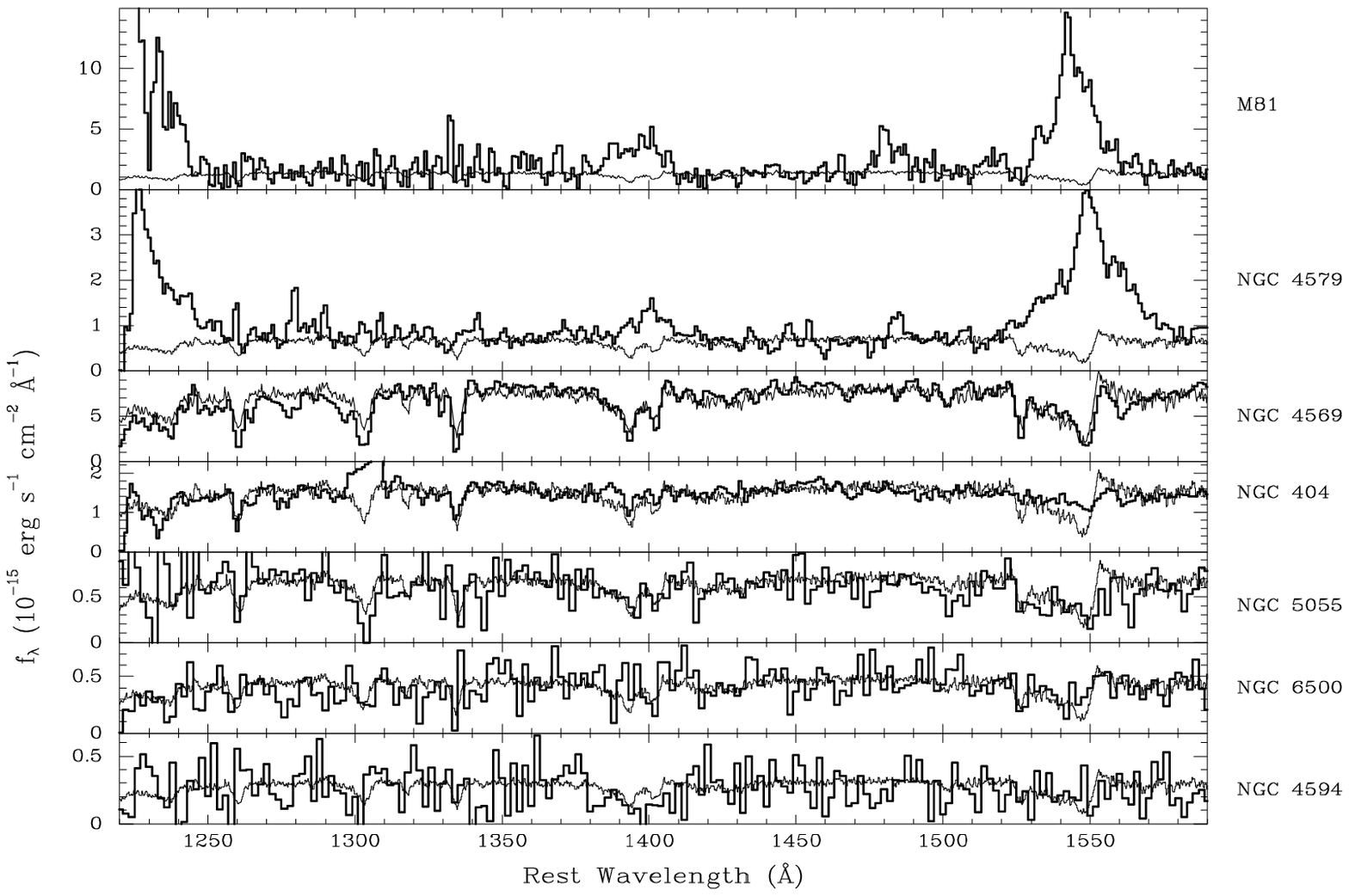,height=3.8truein,width=5.5truein,angle=0}
\hskip 0.1truein
}
}
Fig. 1: FOS G130H spectra of the seven LINERs (bold lines), ordered with the
two broad-lined objects on top, and then with decreasing $f_{\lambda}$.
Overlayed in each case is the spectrum of the
starburst in NGC 1741-B, normalized to be flat in $f_{\lambda}$ and scaled
by a multiplicative factor to match the LINER continuum level.
\vskip +0.3cm
\end{figure}

\begin{figure}
\vbox{
\vbox{
\hskip 0.1truein
\psfig{file=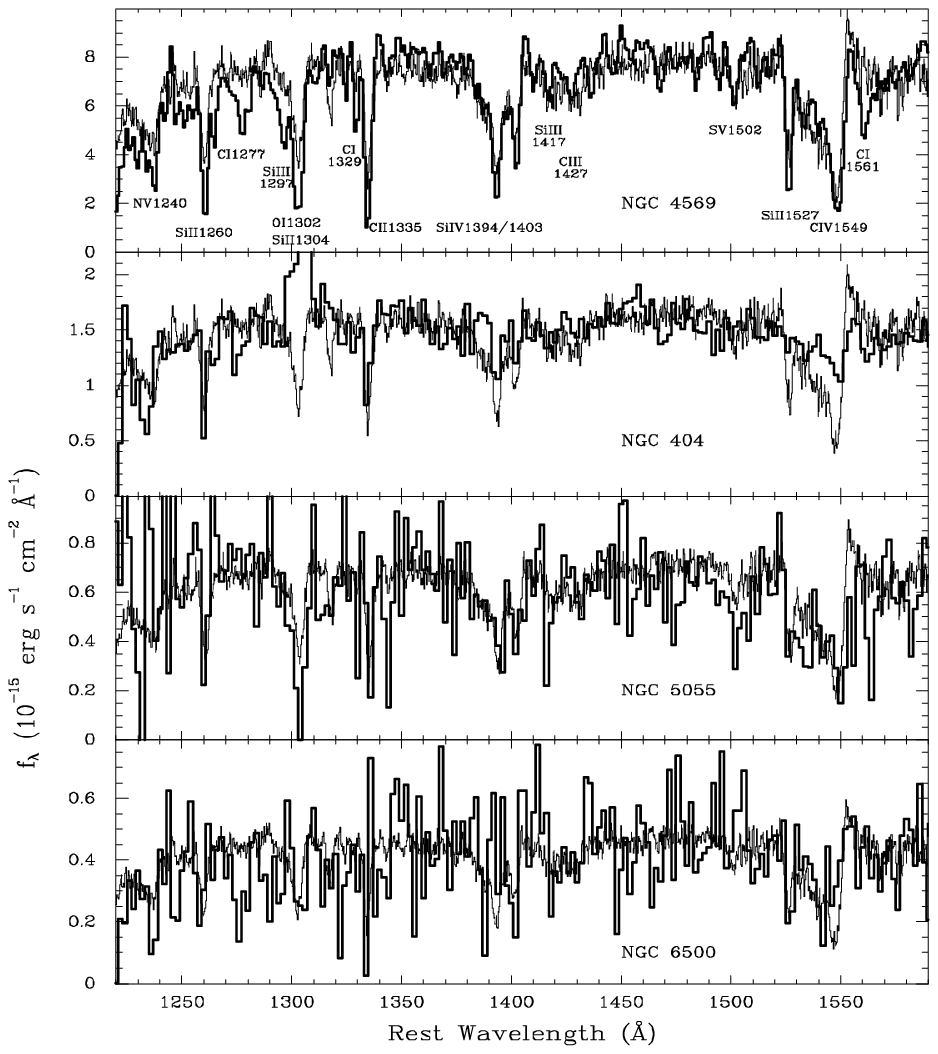,height=6.6truein,width=8.8truein,angle=0}
\hskip 0.1truein
}
}
\vskip +0.3cm
Fig. 2: Same as Figure 1, but only for the four brightest
LINERs devoid of broad emission lines. The strongest absorption
lines are marked for NGC 4569. The emission feature near 1300~\AA\ in
NGC 404 is an artifact.
\end{figure}

The FOS spectrum of NGC 4569, the brightest (and highest S/N)
LINER in our sample, is virtually identical to that of NGC 1741-B.
Most of the features in the two spectra match one-to-one.
NGC 4569 has all the main interstellar absorption lines seen in starburst 
spectra, as well as the broad stellar wind features. Also
detected are several narrow absorption lines which
cannot be interstellar (they are not 
resonance lines) and constitute further evidence for the photospheres 
of hot stars: Si~III $\lambda\lambda 1294/1297$, 
Si~III $\lambda1417$, C III~$\lambda 1427$, S V $\lambda 1502$, and
N IV $\lambda 1720$ (Heckman \& Leitherer 1997).
The correspondence between the two spectra is such that 
a dilution of the stellar features by a nonstellar (i.e., featureless)
 continuum producing more than 20\% of the UV continuum would be
readily apparent.
The only significant difference between the spectra of the starburst
clump in NGC 1741 and the nucleus of NGC~4569 is the presence of
relatively strong 
C I absorption lines (at rest wavelengths 1277~\AA, 1280~\AA, 1329~\AA,
1561~\AA, and 1657~\AA) in NGC 4569.\footnote{
The C I lines are generally not seen with such strength
(up to 3.5 \AA\ equivalent width) in starburst spectra. 
C I has a low
ionization potential (11.26 eV), so it can form only in places that are
shielded from UV radiation by dust, at the interfaces of molecular clouds.
Deeper inside the clouds the carbon gets bound in CO. 
Interestingly, the same
C I lines seem to be present in some of the other LINERs as well, so they
may be a signature of a particular sort of environment.} All the absorption lines 
are at the slight blueshift ($-223$ km s$^{-1}$) of the galaxy,
and hence are not produced in the Milky Way's ISM.
There is no doubt that the UV emission 
in this object, even though its source is highly compact ($\ltorder 2$ pc; Maoz et
al. 1995), is dominated by a cluster of massive stars.
 The luminosity of a star cluster is dominated by
the most massive stars. Using spectral synthesis models
and assuming a distance of 9 Mpc,
we estimate that 250-600 O-type stars (depending on the presence of very massive
early-O-type stars) are sufficient to produce
the observed luminosity. 
A plausible
amount of dust extinction can raise this number by factors of up to
several tens.

Proceeding to the far-UV spectrum of NGC 404 (Fig. 2), we note that 
the broad, blueshifted C~IV absorption is shallower,
but definitely detected.  Si IV absorption
is not seen,
except in the narrow components. According to the synthesis by
 Leitherer et al. (1995), such spectra characterize 
a starburst that is either very young ($<3$ Myr) or $>5$ Myr old. 
The denser winds of O-type supergiants, which are present only
during a limited time, are required in order to produce the broad blueshifted Si IV
absorption. The C~IV absorption, on the other hand, remains as long as 
there are main-sequence O-type stars. The relative shallowness of the C~IV
profile can also be reproduced well by diluting the NGC 1741-B starburst
 spectrum by a featureless (e.g. nonstellar) 
spectrum that is constant in $f_{\lambda}$, and contributes
$\sim 60$\% of the total flux. The strongest interstellar absorption
lines that appear in NGC 4569, including the C I lines, are seen in NGC 404
as well. The velocity width and blueshift of the C~IV profile
clearly point to a stellar origin for this line.
 (The ``emission'' between 1295~\AA\ and 1310~\AA\ in NGC 404 
is an artifact due to a noisy diode
 that masks the O I and S II absorptions that 
are probably present at those wavelengths.) It thus appears that 
the UV continuum source in this
LINER is a star cluster that is of different age than the one in
NGC 4569, or perhaps a 
cluster of similar age whose emission is diluted by a nonstellar continuum.
At a 2 Mpc distance to NGC 404, its UV luminosity
is 100 times lower than that of NGC 4569, implying just two to six 
O-type stars can produce the observed luminosity. Again, the actual number
is likely higher after a reasonable extinction correction.

Continuing to the G130H spectrum of NGC 5055 (Fig. 2), 
the S/N degrades, but 
the blueshifted broad C IV and Si IV 
absorptions seen in NGC 4569 are definitely present.
Their minima are at the galaxy's redshift  ($497$ km s$^{-1}$).
The stellar nature of this UV source is not  surprising,
since Maoz et al. (1995) already noted that it is
marginally resolved at 2200 \AA, with FWHM$\approx 0.2''$ ($\approx 6$ pc).
NGC 4569 and NGC 5055 are remarkably similar
over the entire UV range.
Overall, the spectrum of NGC~5055 resembles a lower S/N version
of that of NGC~4569.
In the spectrum of NGC 6500 (Fig. 2), 
the blueshifted broad C IV absorption is
possibly recognizable although, as noted by Barth et al. (1997), its
significance is arguable when the spectrum is viewed individually.
 The comparison to the spectra of the
other LINERs and the starburst spectrum,
 combined with the obvious degradation in S/N in
this fainter object, suggest that the absorption may, in fact, be present.
The spectrum is too noisy and dominated by scattered light
to reach any conclusion regarding the Si IV absorption. As noted
above, Barth et al. (1997) already concluded, based on the
extended ($0.5''\approx 100$ pc) appearance of the UV source in a WFPC2
image, that this source is probably stellar in nature.
Finally, the NGC 4594 G130H data (Fig. 1) are too noisy and dominated
by scattered light to reach any conclusion, except that
they could be consistent with the same type of spectrum.
There are certainly no strong {\it emission} lines in this part 
of the spectrum of NGC~4594.

We conclude that, in the five LINERs without  
emission lines in the G130H range, the UV continuum is, with varying degrees
of certainty, produced by massive, young stars. 
Could the same be true of the two broad-lined LINERs
in our sample, M81 and NGC 4579? As already noted by Ho et al.
(1996) for M81, one cannot answer this question based on these
UV data alone. Figure 1 shows that if the continuum spectrum
were of the same starburst type as in the other LINERs, we 
would not know it because
the broad emission lines are coincident with the broad
absorptions. 
A case in point
is the recent study by Heckman et al. (1997) of the Seyfert 2
galaxy Mrk 477. The relatively narrow emission lines are
superposed on the broad stellar-wind signatures, but in Si IV and N V
enough of the stellar absorptions are visible near the blue wings
of the emission lines to reveal the starburst nature of the UV continuum
emission.  In M81 and NGC 4579, however, the
emission lines are too broad to see the stellar absorptions, if
they are there.
The single argument that the UV continuum source in NGC 4579 is 
necessarily nonstellar is its  possible factor $\sim 3$ variability reported in  
Barth et al. (1996b).

Barring perhaps
NGC 4579 (assuming its UV variability is real), the UV
continuum in {\it all} the LINERs could be stellar in origin.
Note that this result does not bear directly on the question
of whether or not there is also a nonstellar, quasar-like
object in the nucleus. A microquasar could still
be present, and dominate the emission at wavelengths other
than the UV. Furthermore, it is becoming
increasingly appreciated that circumnuclear
starbursts, albeit on physical scales larger than those
considered here, can contribute 
significantly to the UV and optical brightness of some
AGNs (e.g., IC 5135 -- Shields \& Filippenko 1990; Heckman et al. 
1998; NGC 1068 -- Thatte et al. 1997; NGC 7469 --
Genzel et al. 1995; Mrk 477 -- Heckman et al. 1997; 
some Seyfert 2s -- Gonzalez-Delgado et al. 1998).

\section{THE IONIZING PHOTON BUDGET}

We have compiled \Ha fluxes for each object from Ho et al. (1997a),
Stauffer (1982; NGC 5055), and Keel (1983; NGC 4569).
Figure 3 (left panel) shows the \Ha flux  
vs. $f_{\lambda}(1300~{\rm\AA})$
measured from the spectra and corrected for Galactic
extinction.
There appears to be little relation
between these observables. 
We have computed the ratio of the H$\alpha$ line flux to the
the 1300~\AA\ continuum flux density for young star clusters
containing populations of O-type stars. 
Further details of these and related computations are described
in
Maoz et al. (1998) and Sternberg (1998). Here we consider young clusters
($\ltorder 10^6$ yr old) with Salpeter initial-mass functions (IMFs).
Assuming case-B recombination in $10^4$ K ionization-bounded nebulae, the 
H$\alpha$ line luminosities per number of O-type stars are equal
to 1.4$\times 10^{36}$ and 1.4$\times 10^{37}$ erg s$^{-1}$ for
IMFs which extend up to 30 $M_\odot$ and 120 $M_\odot$, respectively.
For such clusters the H$\alpha/f_{\lambda}(1300$~\AA) ratios equal about 10.5 and 42.0 \AA,
and decrease with increasing cluster age.
 The stellar-wind signatures in several of the
LINERs require the presence of stars of mass $\gtorder 30 M_{\odot}$.
 The two solid diagonal lines in Figure 3 show the maximum \Ha flux
that can be produced with 100\% covering
factor from
 ionization by a stellar population with the given 1300 \AA\ flux
and upper mass cutoff of $120 M_{\odot}$ or $30 M_{\odot}$. 
The two dotted lines show this limit for ionization by power-law continua
of the form $\lambda^{\beta}$ with $\beta=-1$ or $\beta=0$.

\begin{figure}
\vbox{
\vbox{
\hskip -0.5truein
\psfig{file=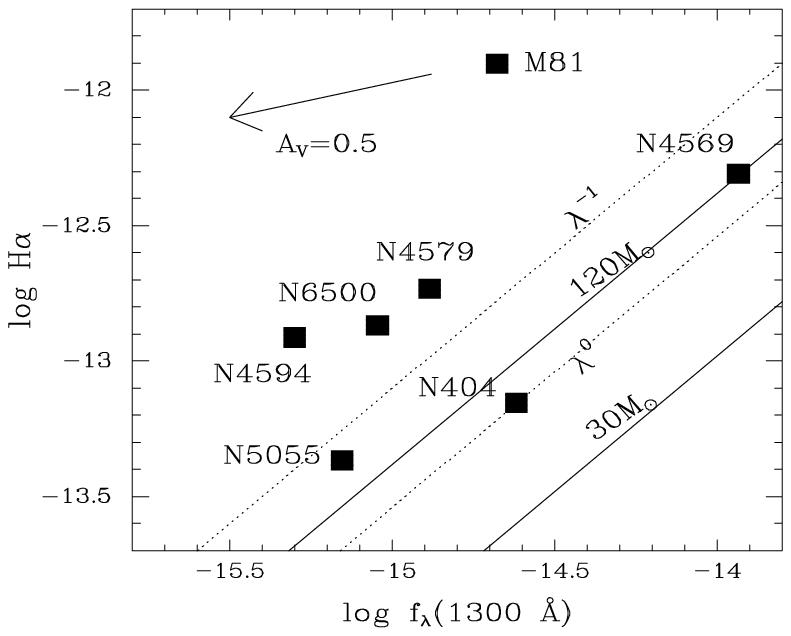,height=4.2truein,width=4.4truein,angle=0}
\hskip -2truein
\psfig{file=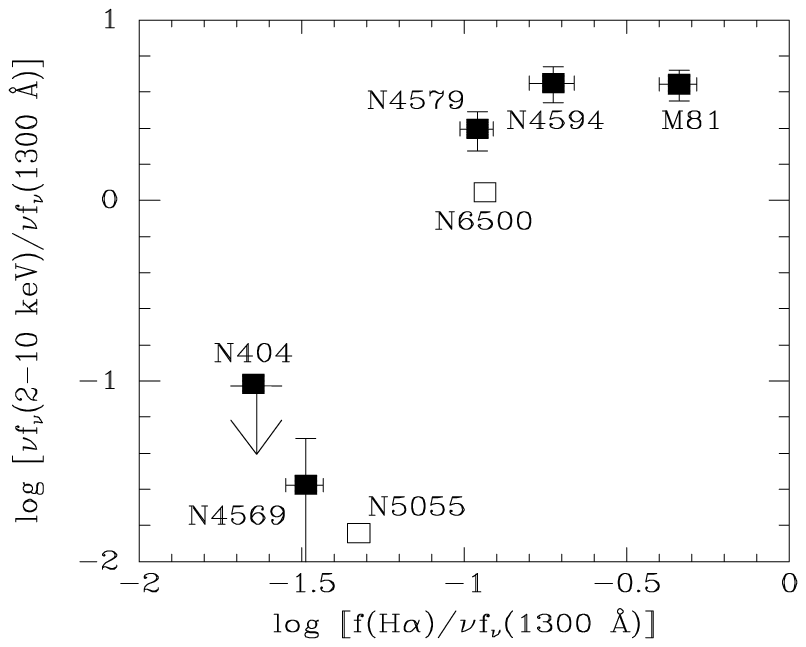,height=4.2truein,width=4.4truein,angle=0}
}
}
\vskip +0.cm
Fig3 : {\bf Left}: Log \Ha flux (in 
erg s$^{-1}$ cm$^{-2}$)  
vs. log $f_{\lambda}(1300{\rm \AA})$ (in 
erg s$^{-1}$ cm$^{-2}$ \AA$^{-1}$)
measured from the spectra and corrected for Galactic
extinction.
The two solid diagonal lines show the maximum \Ha flux
that can be produced in Case B recombination with 100\% covering
factor from
 ionization by a stellar population with a given 1300 \AA\ flux,
resulting from an instantaneous burst of 
 age 1 Myr, Salpeter initial-mass
function, and upper mass cutoff of $120 M_{\odot}$ or $30 M_{\odot}$. 
The two dotted lines show this limit for ionization by power-law continua
$f_{\lambda}\propto\lambda^{\beta}$ with $\beta=-1$ or $\beta=0$.
An  $A_V=0.5$ mag foreground extinction vector, assuming a Galactic
extinction curve, is shown for reference. Typical uncertainties
are 10\% in $f_{\lambda}(1300~{\rm\AA})$ ($\pm 0.04$ in the log) and 
30\% in \Ha flux ($\sim \pm 0.13$ in the log).
{\bf Right}: The ratio of X-ray-to-UV power vs. 
H$\alpha$-to-UV power for each of the LINERs.
Filled symbols denote {\it ASCA} measurements, empty
symbols are extrapolated {\it ROSAT} values.
 Note how in the
four ``UV-photon starved'' objects (high H$\alpha$-to-UV power
ratio) the X-ray power is comparable to, or greater than,
the UV power. An energy source emitting
primarily in the X-rays, and not necessarily related to the
observed UV source, is probably the main ionizing agent in the
four ``UV-photon starved'' objects.
\end{figure}

 We see that, if there is no internal extinction, then some
and possibly all of the LINERs
require an additional ionizing source to drive their line flux.
M81, NGC 4594, and NGC 6500, even if their UV spectra are extrapolated as power
laws with $\beta=-1$ (i.e., harder than observed) rather than as 
stellar-population spectra, also have a severe ionizing photon
deficit. NGC 4579, on the other hand, does have a $\beta\approx -1$ slope.
 If the Maoz et al. (1995) flux
level is adopted, then as before, it has a factor 2.8 ionizing
photon surplus, rather than a deficit. If, however, stars dominate
the UV continuum or the FOS flux level is the ``normal'' level,
then this LINER also has an ionizing photon deficit. In the three LINERs
whose UV emission is clearly dominated by stars (NGC 404, NGC 4569,
and NGC 5055), ionization by the stellar population
can provide the required power, but only if very massive stars
are still present.
 Interestingly, it is these three objects which
also have the lowest emission-line ratios
of [O~III]~$\lambda$5007/H$\beta$, [O~I]~$\lambda$ 6300/H$\alpha$, 
and [S~II] $\lambda\lambda$6716, 6731/\Ha in the sample (see Ho et al. 1997a); 
this is exactly what one
would expect from ionization by the relatively soft continuum of a
significant stellar component, which produces less heating per ionization
and a smaller partially ionized zone than an AGN-like power-law
continuum.

Alternatively, some internal foreground extinction (such that
the UV emission is attenuated as seen by the observer, but not 
as seen by the ionized gas) could, in principle, alleviate
the ionization budget problem in some of the objects.
The uncertainty in the extinction
curve and the many possible configurations for different extinction
of the continuum and nebular emission result in a broad
range of possible extinction corrections.
To illustrate the effect of a plausible extinction correction, 
Figure 3 shows an  $A_V=0.5$ mag foreground
extinction vector, assuming a Galactic
extinction curve. A Galactic curve is intermediate in ``greyness''
to the SMC and Calzetti et al. (1994) curves.
Note that the objects with the most severe ionizing photon
deficits (M81, NGC 4594, and NGC 6500) 
are those whose Balmer decrements  indicate
little internal reddening, $A_V=0.1$ to 0.35 mag, for the range
in extinction curves.
In NGC 4569, NGC 5055, and
NGC 4579, a plausible extinction correction
will increase the intrinsic
 1300~\AA\ flux by a factor $\gtorder 10$ (with
 a corresponding increase in the number
of O-type stars),
and may relieve the need for the most massive
 ($\sim 120 M_{\odot}$) stars.

 We conclude that at least some
of the LINERs in our sample have an ionizing photon
deficit, indicating an additional energy source, beyond
that implied by the observable UV.
To search for evidence for such an additional 
source, we have compiled X-ray data for the seven LINERs.
For NGC 404, NGC 4569, NGC 4579, and NGC 4594 we have
used analysis of both archival and new {\it ASCA} data by
Y. Terashima, A. Ptak, and L. Ho (see Terashima, this volume) to
obtain unabsorbed 2--10 keV fluxes. For M81 we have used the
{\it ASCA} flux derived by Ishisaki et al. (1996). 
NGC 5055 and NGC 6500 have not been observed with {\it ASCA},
but have been observed by {\it ROSAT} in the 0.1--2.4 keV
band. We have extrapolated these measurements
to the {\it ASCA} bandpass, as described in Maoz et al. (1998),
though of course such extrapolations are highly uncertain
and should be treated with caution.

Figure 3 (right panel) shows the ratio of X-ray-to-UV power vs. 
H$\alpha$-to-UV power for each of the LINERs. Note how in the
four ``UV-photon starved'' objects (high \Ha to UV power
ratio) the X-ray power is comparable to or greater than
the UV power. Conversely, in the three objects without
a serious ionizing photon deficit, the X-ray power is one to two
orders of magnitude lower than the UV power.
This suggests that, indeed, an energy source emitting
primarily in the extreme-UV, and not directly related to the
observed UV source, may be the main ionizing agent in the
four ``UV-photon starved'' objects. The observed X-ray emission
would be the high-energy tail of such a component.
For example, a blackbody with temperature $>3.3\times 10^5 {\rm K}$ 
would have
an X-ray to UV power ratio greater than or equal to that of the
UV-photon starved objects. Alternatively, the photoionizing continua in
 these objects could consist
of power-law spectra extending from the X-rays to the Lyman limit.
However, this would require that the central UV sources 
are significantly attenuated by dust extinction while the nebular 
components are not.

It has long been debated whether the emission lines in
LINERs are produced by means of photoionization or shocks.
The question has been recently re-addressed in the analysis
of the four LINERs among the seven discussed here with
published UV spectra (Ho et al. 1996; Barth et al. 1996b, 1997;
Nicholson et al. 1998). These studies show that the UV line
ratios are consistent with either photoionization by an
AGN-like continuum or by slow-moving shocks, but inconsistent
with the fast ``photoionizing'' shocks proposed by Dopita \& Sutherland (1996).
The data discussed here show that
massive stars exist in some LINERs. The sources
required to produce most or all of the emission lines by
photoionization are therefore present, at least sometimes.  

\newpage
\section{CONCLUSIONS}

The \hst UV spectra of seven LINERs having
compact nuclear UV sources reveal the following.\\
1. At least three of the LINERs have clear spectral signatures
indicating that the dominant UV continuum source is a cluster of massive stars.\\
2. A similar continuum source could dominate in the other four
LINERs as well, but its spectral signatures would be veiled by
low S/N or superposed broad emission lines. 
Alternatively, there may be two types of UV-bright LINERs: those 
where the UV continuum is produced by a starburst, and those where
it is nonstellar. If the variability
of NGC 4579 is real, its continuum source is certainly an AGN.\\
3. The three ``clearly-stellar'' LINERs
have relatively weak X-ray emission, and their stellar populations probably
provide enough ionizing photons to explain the observed optical
emission-line flux. The four other LINERs have severe ionizing
photon deficits, for reasonable extrapolations of
 their UV spectra beyond the Lyman limit, but an X-ray/UV power ratio
that is higher by two orders of magnitudes than that of the
three stellar LINERs. A component
which emits primarily in the extreme-UV may be the main 
photoionizing agent in these four objects. 

The picture emerging from this comparison is that the compact 
UV continuum source seen in $\sim 25\%$ of LINERs (Maoz et al. 1995;
Barth et al. 1996a, 1998) is, at least in some cases, a nuclear 
starburst rather than an AGN-like nonstellar object. The UV luminosity
is driven by tens to thousands of O-type stars, depending on
the object and the extinction assumed. The O-stars could be the high-mass
end of a bound stellar population, similar to those seen in
super star clusters (e.g., Maoz et al. 1996b) .
 Nonstellar
sources in LINERs may be significant or even dominate at other 
wavelengths, as we indeed find for some of the objects. Even
the three ``clearly-stellar'' LINERs, which do not obviously require
the existence of such an additional source, may well have one; the
ionizing photon budget estimate was made assuming a 100\% covering
factor of the line-emitting gas, which is not necessarily true.
This picture fits well with recent results showing that nuclear-starburst
and quasar-like activity are often intermingled in Seyfert 1 and 2
galaxies. For example, Gonzalez-Delgado et al. (1998) show that
the so-called ``featureless continuum'' that has been previously observed
in some Seyfert 2s is not really featureless, and is due to
massive stars. Their preliminary analysis of the 20 brightest
Seyfert 2s indicates that massive stars are present in at least
one-third of the cases. Our results extend this result to the lower luminosities
of the LINERs discussed here, although the question of whether a 
``micro-quasar'' exists in these objects is still open.

Higher S/N UV spectra of some of the objects in our sample could reveal
whether they too have the signatures of massive stars. Conversely,
further evidence of UV continuum variability, as suggested in NGC 4579,
should be sought in LINERs. Even in those LINERs with stellar signatures,
UV variability could reveal a contribution by a nonstellar component.
This work suggests, however, that the AGNs possibly associated with LINERs 
are most prominent at higher energies.
X-ray observations by upcoming space missions, having better angular and spectral
resolution and higher sensitivity, will likely provide key insights
to the nature of LINERs.

\acknow
I thank the organizers for a stimulating and enjoyable meeting.


\begin{references}

\noindent\\ Barth, A.J., Ho, L.C., Filippenko, A.V., \& Sargent, W.L.W.
1996a, in ``The Physics of LINERs in View of Recent Observations'', eds.
M. Eracleous et al. (San Francisco: ASP), p. 153
\\ Barth, A.J, Reichert, G.A., Filippenko, A.V., 
Ho, L.C., Shields, J.C., Mushotzky, R.F.
 \& Puchnarewicz, E.M. 1996b, AJ, 112, 1829
\\ Barth, A.J, Reichert, G.A., Ho, L.C., Shields, J.C., 
Filippenko, A.V., \& Puchnarewicz, E.M. 1997, AJ, 114, 2313
\\ Barth, A.J., Ho, L.C., Filippenko, A.V., \& Sargent, W.L.W.
 1998, ApJ, 496, 133
\\ Calzetti, D., Kinney, A.L., \& Storchi-Bergmann, T. 1994, ApJ, 429, 582
\\ Conti, P.S., Leitherer, C., \& Vacca, W.D. 1996, ApJ, 461, L87
\\ Devereux, N., Ford, H., \& Jacoby, G. 1997, ApJ, 481, L71
\\ Dey, A., van Breugel, W., Vacca, W.D., \& 
    Antonucci, R. 1997, ApJ, 490, 698
\\ Filippenko, A. V., \& Sargent, W. L. W. 1985, ApJS, 57, 503
\\ Filippenko, A. V., \& Sargent, W. L. W. 1988, ApJ , 324, 134
\\Gonzalez-Delgado, R.M., et al. 1998, ApJ, in press
\\ Heckman, T.M. 1980, A\&A, 87, 152
\\ Heckman, T.M., \& Leitherer,  C. 1997, AJ, 114, 69
\\ Heckman, T.M., et al. 1997, ApJ, 482, 114
\\ Ho, L.C., Filippenko, A.V., \& Sargent, W.L.W. 1996, ApJ, 462, 183
\\ Ho, L. C., Filippenko, A. V., \& Sargent, W. L. W. 1997a,
ApJS, 112, 315
\\ Ho, L. C., Filippenko, A. V., Sargent, W. L. W., \& Peng, C. Y.
 1997b , ApJS, 112, 391
\\ Keel, W.C. 1983, ApJ, 269, 466
\\ Leitherer, C., \& Heckman, T.M. 1995, ApJS, 96, 9
\\ Leitherer, C., Vacca, W.D., Conti, P.S., Filippenko, A.V.,
Robert, C., \& Sargent, W.L.W. 1996, ApJ, 465, 717
\\  Leitherer, C., Robert, C., \& Heckman, T.M. 1995, ApJS, 99, 173
\\ Lowenthal, J.D., et al. 1997, ApJ, 481, 673
\\ Maoz, D., Barth, A.J., Sternberg, A., Filippenko, A.V.,
Ho, L.C., Macchetto, F.D., Rix, H.-W., \&  Schneider, D.P. 1996b, AJ, 111, 2248
\\ Maoz, D., Filippenko, A.V.,
Ho, L.C., Macchetto, F.D., Rix, H.-W., \&  Schneider, D.P. 1996a, ApJS,
107, 215
\\ Maoz, D., Filippenko, A. V., Ho, L. C., Rix, H. -W.,
Bahcall, J. N., Schneider, D. P., \& Macchetto, F. D. 1995, ApJ, 440, 91
\\ Maoz, D., Koratkar, A.~P., Shields, J.~C., Ho, L.~C., Filippenko, A.~V., \&
Sternberg, A. 1998, AJ, 116, 55
\\ Nicholson, K.L., Reichert, G.A., Mason , K.O., Puchnarewicz, E.M.,
Ho, L.C., Shields, J.C., \& Filippenko, A.V. 1998, MNRAS, submitted
\\ Shields, J. C., \& Filippenko, A. V. 1990, AJ, 100, 1034
\\ Stauffer, J.R. 1982, ApJ, 262, 66
\\ Steidel, C.C., Giavalisco, M., Dickinson, M., \& Adelberger, K.L.
  1996, AJ, 112, 352
\\ Sternberg, A. 1998, \apj, in press
\\ Thatte, N., Quirrenbach, A., Genzel, R., Maiolino R., \& Tecza, M.
1997, ApJ, 490, 238
\\ Vacca, W.D., Robert, C., Leitherer, C., 
   \& Conti, P.S. 1995, ApJ, 444, 647

\end{references}
\end{document}